\documentclass[10pt,preprint,nofootinbib]{revtex4-2}
\usepackage{graphicx}
\usepackage{hyperref}
\usepackage{color}
\newcommand{\be}{\begin{equation}}
\newcommand{\ee}{\end{equation}}
\newcommand{\beq}{\begin{eqnarray}}
\newcommand{\eeq}{\end{eqnarray}}
\newcommand{\ba}{\begin{array}}
\newcommand{\ea}{\end{array}}

\begin{document}

\title{JLab and J-PARC for the J/{\ensuremath{\psi}} Production at the Threshold}

\date{\today}

\author{\mbox{Igor~I.~Strakovsky}}
\altaffiliation{Corresponding author: \texttt{igor@gwu.edu}}
\affiliation{Institute for Nuclear Studies, Department of Physics, The 
George Washington University, Washington, DC 20052, USA}

\author{\mbox{Jung~Keun~Ahn}}
\altaffiliation{Corresponding author: \texttt{ahnjk@korea.ac.kr}}
\affiliation{Department of Physics, Korea University
145 Anam-ro, Seongbuk-gu, Seoul 02841, Republic of Korea}

\author{\mbox{William~J.~Briscoe}}
\altaffiliation{Corresponding author: \texttt{briscoe@gwu.edu}}
\affiliation{Institute for Nuclear Studies, Department of Physics, The 
George Washington University, Washington, DC 20052, USA}

\author{\mbox{Misha~G.~Ryskin}}
\altaffiliation{Corresponding author: \texttt{ryskinmg@airnet.ru}}
\affiliation{NRC ``Kurchatov Institute'', Petersburg Nuclear Physics 
Institute, Gatchina, 188300, Russia}

\author{\mbox{Axel~Schmidt}}
\altaffiliation{Corresponding author: \texttt{axelschmidt@gwu.edu}}
\affiliation{Institute for Nuclear Studies, Department of Physics, The 
George Washington University, Washington, DC 20052, USA}

\noaffiliation

\begin{abstract}
New threshold measurements for $\gamma~p\to p~J/\psi \to p~(\mu^+\mu^-)$ by 007 and 
$\gamma^\ast~p\to p~J/\psi \to p~(e^+e^-)$ by CLAS12 allow us to extend the previous 
phenomenological determination of the J/\ensuremath{\psi}-proton scattering length, 
$\alpha_{J/\psi p}$, using GlueX threshold data for $\gamma~p\to p~J/\psi \to p~
(e^+e^-)$. The agreement between all three J/\ensuremath{\psi} data sets shows no 
indication of systematic differences between methodologies. Furthermore, perturbative 
QCD predictions support the phenomenological determination of heavy vector meson-nucleon scattering lengths. The role of the nucleon form factor, 
$F_N$, is discussed, and a possible correction to the phenomenological scattering 
length for heavy vector meson photoproduction is calculated using the pole form of 
$F_N$. Upcoming J-PARC threshold measurements of the reaction $\pi^-~p\to n~J/\psi \to 
n~(l^+l^-)$ will help to evaluate the possible role of heavy pentaquark, $P_c$, states 
in low-energy $J/\psi$ production and the effects caused by the nucleon form factors. 
\end{abstract}

\maketitle

\section{Introduction}\label{sec:sec1} 
There are no vector meson beams, so modern experiments using electromagnetic (EM) facilities attempt to access vector meson (V)–nucleon (N) interactions via EM production reactions, such as $e~p \to e^\prime~V~p$ or $\gamma~p\to V~p$. The J/{\ensuremath{\psi}} is a member of the vector meson family and has the same spin ($J$), parity ($P$), and charge conjugation ($C$) quantum numbers as the photon: $J^{PC} = 1^{-~-}$. One may therefore apply Sakurai’s vector meson dominance (VMD) model~\cite{Sakurai:1960ju}, assuming that a real photon can fluctuate into a virtual $V$ (J/\ensuremath{\psi} in our case), which subsequently scatters off a target nucleon.

The discovery of the J/\ensuremath{\psi}, the first heavy vector meson, 50 years ago at BNL~\cite{E598:1974sol} and SLAC~\cite{SLAC-SP-017:1974ind} was termed the ``November Revolution.'' Beyond confirming the existence of the charm ($c$) quark and opening access to the study of the $c\bar{c}$ system, it has allowed the exploration of a variety of interesting aspects of physics. In particular, the heavy quark mass dependence of the properties of systems containing the heavy quark $Q$. Recall that the $Q\bar{Q}$ system is expected to have a small size $\propto 1/m_Q$ and that the predictions of perturbative QCD (pQCD) theory should be better justified here.

The Thomas Jefferson National Accelerator Facility (JLab), which currently operates with a $12~\mathrm{GeV}$ beam of electrons, has the capacity to measure J/{\ensuremath{\psi}} photoproduction cross sections at threshold with high statistics. The threshold energy for a fixed proton target is $E = 8.2~\mathrm{GeV}$. Table~\ref{tbl:tab1} lists several JLab measurements reported recently. The behavior of the cross section at threshold is sensitive to the scattering length, $\alpha_{J/\psi p}$, of the J/{\ensuremath{\psi}}-proton system.
\begin{table}[htb!]
\centering \protect\caption{JLab measurements of the J/\ensuremath{\psi} photoproduction on the proton target at the threshold.
1st column is the list of experiments,
2nd column is the luminosity for them, and
3rd column is the number of J/\ensuremath{\psi} events above the threshold.
}
\vspace{2mm}
{%
\begin{tabular}{|c|c|c|}
\hline
Experiment                  & Luminosity        & $J/\psi$ \tabularnewline
                            & (pb$^{-1}$)       & (events) \tabularnewline
\hline
GlueX~\cite{GlueX:2019mkq}  &  68               & 469$\pm$22 \tabularnewline
GlueX~\cite{GlueX:2023pev}  & 320               & 2270$\pm$58 \tabularnewline
007~\cite{Duran:2022xag}    & $8.68\times 10^6$ & $\sim$2000 \tabularnewline
007~\cite{007:2026dow}      & $8.68\times 10^6$ & 1041$\pm$55 \tabularnewline
CLAS12~\cite{Chatagnon:2026qsv}& 149            & 780$\pm$40 \tabularnewline
\hline
\end{tabular}} \label{tbl:tab1}
\end{table}

The JLab measurements of J/{\ensuremath{\psi}} production are an important link in a chain of vector-meson photoproduction measurements that enable a phenomenological analysis of trends in the vector-meson-proton scattering length (SL). Previous analyzes include: $\gamma~p\to \rho~p$ (from CLAS12)~\cite{Strakovsky:2025ews, Wang:2022zwz}, $\gamma~p\to \omega~p$ (from A2 at MAMI, ELPH, and CBELSA)~\cite{Strakovsky:2014wja, Ishikawa:2019rvz, Han:2022khg}, $\gamma~p \to \phi~p$ (from CLAS6 and LEPS)~\cite{Strakovsky:2020uqs, Han:2022khg}, and $\gamma~p \to J/\psi~p$ (from GlueX, 007, and CLAS12)~\cite{Strakovsky:2019bev, Strakovsky:2025rsm}. While there has not yet been a measurement of threshold $\Upsilon$-meson photoproduction, this will be possible at the future Electron Ion Collider (EIC), enabling an extension of the analysis to the $\Upsilon$-proton system~\cite{Strakovsky:2021vyk}.  Results of these studies are shown in Table~\ref{tbl:tab2}. 

Note that the SLs listed in Table~\ref{tbl:tab2} are much smaller than the size of either the proton ($\sim 1~\mathrm{fm}$) or the vector meson. This may be explained by the ``young’’ meson hypothesis proposed first in~\cite{Feinberg:1980yu} (see~\cite{Strakovsky:2021vyk} for a discussion of the phenomenological treatment).

The point is that in photoproduction, the $Q\bar{Q}$ pair is produced at that 
point (locally) by the point-like photon, and the heavy quarks do not have 
sufficient time to fly away at the distances corresponding to the normal 
size of the vector meson wave function. That is, actually, we deal with the small 
size $Q\bar{Q}$ pair-proton interaction and not with the normal $V$-proton 
scattering. On the other hand, within pQCD, the corresponding 
cross section $\sigma((Q\bar{Q})+p)\propto r^2$, where $r$ is the antiquark-quark separation. Indeed, the small size (young) colorless object is almost 
sterile with respect to the QCD interaction. Therefore, the scattering length, 
\textit{i.e.}, the cross section measured in photo(electro)production, is smaller than that for the $Vp\to Vp$ interaction. This is an interesting effect 
(sometimes called the puzzle) that we can observe in photoproduction 
experiments.
\begin{table*}[htb]
\caption{Vector Meson Photoproduction for reactions $\gamma~p \to V~p$, where V = \ensuremath{\rho}, \ensuremath{\omega}, \ensuremath{\phi}, J/\ensuremath{\psi}, and \ensuremath{\Upsilon}. 
1st column is the name of the Vs.
2nd column is their masses~\cite{ParticleDataGroup:2024cfk}.
3rd column is the photon threshold momentum in the center-of-mass (CM) for V photoproduction on the proton.
4th column is the partial decay width $\Gamma(V \to 
e^+~e^-)$~\cite{ParticleDataGroup:2024cfk}.
The 5th column shows the CM energy range for the total cross-section measurements.
6th column is the minimal momentum $q_{min}$ in CM for the V accessible in the photoproduction experiments.
7th column is the Vp SL using a phenomenological treatment by 
Eq.~(\ref{eq:eq4}). The determination for the $\Upsilon$ is a theoretical projection made for the future EIC~\cite{Strakovsky:2021vyk}, and is not based on measured data.}

\begin{center}
\begin{tabular}{|c|ccc|cc|c|}
 \hline
Meson     &Mass    &$k_{thr}$&$\Gamma(V\to e^+~e^-)$&$W_{min}-W_{max}$              &$q_{min}$& $|\alpha_{Vp}|$\tabularnewline
          &(MeV)   & (MeV/c) & (keV) & (MeV)  & (MeV/c) & (fm) \tabularnewline
 \hline
$\rho$    &775.26  & 599.9   & 7.04               & 1737--1987~\cite{CLAS:2018drk}&143      & 0.23$\pm$0.03~\cite{Strakovsky:2025ews} \tabularnewline
          &        &          &  &1873--2017~\cite{Klein:1996} &        & 0.24$\pm$0.02~\cite{Wang:2022zwz} \tabularnewline
$\omega$  &782.66  & 604.7   & 0.60               & 1724--1872~\cite{Strakovsky:2014wja}               &49     & 0.82$\pm$0.03~\cite{Strakovsky:2014wja} \tabularnewline
          &        &         &    & 1710--1740~\cite{Ishikawa:2019rvz}  & &0.97$\pm$0.26~\cite{Ishikawa:2019rvz} \tabularnewline
          &        &         &    & 1770--1970~\cite{CBELSATAPS:2015wwn}                &       & 0.811$\pm$0.019~\cite{Han:2022khg} \tabularnewline
$\phi$   &1019.460& 754.0   & 1.27               &2005-2245~\cite{Dey:2014tfa}          &216    & $0.063\pm 0.010$~\cite{Strakovsky:2020uqs} \tabularnewline
         &            &       &                  & 1922--2320~\cite{Chang:2007fc}  &
         &$0.109\pm 0.008$~\cite{Han:2022khg} \tabularnewline
$J/\psi$  &3096.900& 1908.5   & 5.53             &4054--4709~\cite{GlueX:2019mkq}         &241    & $(3.08\pm 0.55)\times 10^{-3}$~\cite{Strakovsky:2019bev} \tabularnewline
         &        &           &                    &4075--4763~\cite{GlueX:2023fcq}  &166    & $(2.98\pm 0.25)\times 10^{-3}$~\cite{Strakovsky:2025rsm} \tabularnewline
         &         &         & &4254--4542~\cite{Duran:2022xag}  & 588  & \tabularnewline
         &         &         & &4254--4542~\cite{007:2026dow}    & 588  & \tabularnewline
         &         &         & &4085--4542~\cite{Chatagnon:2026qsv} & 286 & \tabularnewline
$\Upsilon$&9460.40 & 5157.0    & 1.340              &10548--12348~\cite{Guo:2021ibg}       &521    & $(0.51\pm 0.03)\times 10^{-3}$~\cite{Strakovsky:2021vyk} \tabularnewline
 \hline
\end{tabular}
\end{center} \label{tbl:tab2}
 \end{table*}

Let us introduce several recent theoretical and Lattice results to compare with the phenomenology.

Tang \textit{et al.} (referred to hereafter as the Nanjing University group) have performed calculations of vector meson photoproduction from threshold to very high energies, assuming a production mechanism comprising a quark loop combined with Pomeron exchange~\cite{Tang:2025qqe}. They extract scattering lengths without requiring a fit to the data, arriving at $|\alpha_{\rho  N}| = 184~\mathrm{mfm}$, $|\alpha_{\phi N}| = 123~\mathrm{mfm}$, $|\alpha_{J/\psi N}| = 6.34~\mathrm{mfm}$, and $|\alpha_{\Upsilon N}| = 0.196~\mathrm{mfm}$. These results are consistent within uncertainties (experimental and theoretical) with the phenomenological results (Table~\ref{tbl:tab2}).

A coupled-channel mechanism with VMD by Du \textit{et al.} (hereafter referred to as the Bonn University group) gives $|\alpha^{J = 1/2}| = 0.2 \ldots 3.1~\mathrm{mfm}$ and $|\alpha^{J = 3/2}| = 0.2 \ldots 3.0~\mathrm{mfm}$, where $J$ corresponds to the total angular momentum of the $J/\psi$p system~\cite{Du:2020bqj}. This is consistent with the phenomenological results as well (Table~\ref{tbl:tab2}).

Wu \textit{et al.} (hereafter referred to as the Tsinghua University group) report a calculation assuming J/\ensuremath{\psi}-nucleon scattering can occur through two distinct mechanisms: the coupled-channel mechanism via open-charm meson-baryon intermediate states ($\alpha_{J/\psi N} = -10 \ldots -0.1~\mathrm{mfm}$) and the soft-gluon exchange mechanism, which is dominant ($\alpha_{J/\psi N} \lesssim -160~\mathrm{mfm}$)~\cite{Wu:2024xwy}. The model accounts only for pions and kaons in the $t$-channel, neglecting contributions from heavier mesons; thus, the final SL may be slightly larger. 

The Lattice HAL Collaboration determined the J/\ensuremath{\psi}-nucleon SL using the elastic $(c\bar{c})p\to (c\bar{c})p$ reaction, resulting in $\alpha(^2S_{1/2}) = 380\pm 40^{+0}_{-30}~\mathrm{mfm}$~\cite{Lyu:2024ttm}. The difference with respect to the phenomenology (Table~\ref{tbl:tab2}) may be explained by the fact that this reaction involves a vector meson in equilibrium (the ``old'' meson). 

We need to have already summarized young versus old at this point... 
Furthermore, the Nanjing, Bonn, and Tsinghua groups calculated SLs in the photoproduction reaction $\gamma~p\to V~p$, whereas in Lattice, the calculation considers the elastic $(c\bar{c})p$ reaction. It is puzzling that in the $(c\bar{c}) p$ case, the SL is much larger. The young effect explains this puzzle.

Note that several previous theoretical results for the VN SL determination, including potential approaches and LQCD calculations, have yielded much larger SLs (see, for instance, Refs.~\cite{Appelquist:1978rt, Peskin:1979va, Liu:2008rza, Aoki:2009ji, Assi:2025ysr}). 
Most likely, such a significant SL results from considerable distances in the tail of the Van der Waals potential, which in QCD should be mitigated by confinement~\cite{Dokshitzer:2024}. 

The evaluation of $\gamma~p \to J/\psi~p$ cross sections of the proton, which were reported by 007~\cite{007:2026dow} and CLAS12~\cite{Chatagnon:2026qsv}, is challenging to extend beyond our previous phenomenological determination of J/\ensuremath{\psi}p SL~\cite{Strakovsky:2019bev, Strakovsky:2025rsm}. As we will show in Section~\ref{sec5}, the agreement in $\gamma~p \to J/\psi~p \to (e^+~e^-)~p$, $\gamma^\ast~p \to J/\psi~p \to (e^+~e^-)~p$, and $\gamma~p \to J/\psi~p \to (\mu^+~\mu^-)~p$ total cross sections between independent high-luminosity GlueX, 007, and CLAS12 measurements, and phenomenological analysis allows us to feel comfortable.

Recall that, strictly speaking, to measure the scattering length of $J/\psi$-proton interaction, we have to consider not the $J/\psi$ photoproduction but $J/\psi$ elastic scattering on the proton. In comparison with the elastic process in the case of photoproduction, we have three additional factors that may depend on the vector meson mass.

The point is that the zero mass of the incoming photon is much smaller than the vector meson mass $M_V$. In terms of VMD, the resulting amplitude is proportional to the $\gamma\to V$ transition coupling $f_V(M_V)$, the amplitude probability not to destroy the target proton $F_N(t)$, and the $c\bar c\to J/\psi$ transition form factor $F_\psi(t)$.

We have to emphasize that to produce a heavy ($J/\psi$) meson at the threshold, we need to have a rather large momentum transferred $t$. The value of $t$ may be calculated using the relation
\begin{equation}
    s+t+u=\sum_{i=1}^4 m^2_i = 2m^2_N+M^2_V \>,
    \label{1}
\end{equation}
where $s$, $t$, and $u$ are Mandelstam variables. $m_{photon}=0$.

Since at the threshold in the proton-$J/\psi$ rest frame, both the vector meson and the proton are at rest, the values of $s$, $t$, and $u$ can be expressed via the energy of the incoming proton, E, as $s=m^2_N+M^2_V+2m_NM_V$, $t=2m^2_N-2Em_N$, and $u=m^2_N+M^2_V-2EM_V$. So we get from (\ref{1})
\begin{equation}
    E=\frac{2m^2_N+M^2_V+2m_NM_V}{2(M_V+m_N)} \quad \mbox{and}
    \nonumber
\end{equation}
\begin{equation}
    t=2m^2_N - m_N \frac{2m^2_N+M^2_V+2m_NM_V}{(M_V+m_N)} \>.
    \label{2}
\end{equation}
For the case of $J/\psi$ at threshold $t=-2.2~\mathrm{GeV^2}$.
Thus, the SL extracted via the VMD from the photoproduction of a heavy vector meson will be smaller than that in elastic meson-proton scattering by the factor $F_N(t)F_\psi(t) f_V(M^2_V)$. Note that the ``elastic'' proton form factor $F_N(t)$ is {\em not equal} to the electromagnetic form factor, $F_e(t)$. Its behavior depends on the particular dynamics of interaction. We do not know $F_N(t)$ experimentally, and for our numerical estimate, we will take the pole form
\begin{equation}
    F_N(t)=\frac 1{1-t/\Lambda^2}
    \label{3}
\end{equation}
with $\Lambda^2=0.71~\mathrm{GeV^2}$ (this $F_N(t)=\sqrt{F_e(t)}$ looks 
reasonable, assuming that the two gluons describe the interaction 
exchange).

Since at the beginning the photon produces the $c\bar c$ pair at the point, and the size of the heavy vector boson is much smaller than the proton size, we expect that the factors $f_V$ and $F_\psi$ should be much flatter as a function of $t$. Therefore (keeping in mind the large uncertainty in $F_N$), we take $f_V F_\psi=1$.

Note also that in the models~\cite{Tang:2025qqe, Du:2020bqj}, where the interaction is mediated by the heavy quark loop and the two gluon exchange, these factors are accounted for automatically.

Besides the form factor suppression, we expect a smaller SL due to the ``young effect,'' where the reaction, rather than proceeding through the $V$-proton interaction, proceeds through the interaction of the recently produced $Q\bar Q$ pair, whose size is smaller than that of the fully formed vector meson.

To study both the ``form factor'' and the ``young effect'' in more detail, it would be interesting to measure the electroproduction at $Q^2$ values comparable to $M_{J/\psi}^2$.

Another interesting piece of information may come from the
J-PARC facility, which can measure the near-threshold V production using neutral low- and high-energy pion beams~\cite{Ryu:2024}.


\section{J/{\ensuremath{\psi}} Threshold Measurements using the GlueX Spectrometer}\label{sec2}
The GlueX spectrometer in JLab's Experimental Hall~D has nearly full acceptance~\cite{GlueX:2020idb}, which allows for a straightforward measurement of the total cross section, $\sigma_t$, for the reaction $\gamma~p \rightarrow J/\psi~p \rightarrow (e^+~e^-)~p$, and eliminates the need to extrapolate to low/high four momentum transfer squared $t$. 
The GlueX Collaboration reported results from data collected in 2016--2017 (68 pb$^{-1}$)~\cite{GlueX:2019mkq}, and then updated this result from data collected over 2016--2018 (320 pb$^{-1}$)~\cite{GlueX:2023pev} (see Table~\ref{tbl:tab1}).  J/\ensuremath{\psi} yields were extracted from fits of $M(e^+~e^-)$ distributions, and the Bethe-Heitler (BH) reaction ($1.2 - 2.5~\mathrm{GeV}$) was used for normalization. The GlueX Collaboration reported 18 total cross-sections as a function of photon energy, $E$, and 3 differential cross-sections as a function of $t$ over the energy range $E = 8.2 - 11.4~\mathrm{GeV}$. 

\section{J/{\ensuremath{\psi}} Threshold Measurements using HMS and SHMS Spectrometers}\label{sec3}
The 007 Collaboration experimented in 2019 in JLab's Experimental Hall~C using an untagged photon beam scattering off a proton target, with the High Momentum Spectrometer (HMS) and the Super High Momentum Spectrometer (SHMS) used to detect lepton pairs from J/\ensuremath{\psi} decay~\cite{Ali:2025dan, Duran:2022xag}. The experiment covered an energy range of $E = 9.2 - 10.5~\mathrm{GeV}$. The collaboration first reported 10 differential cross sections as a function of $t$ for the reaction $\gamma~p \rightarrow J/\psi~p \rightarrow (e^+~e^-)~ p$~\cite{Duran:2022xag}. Subsequently, 007 reported 10 samples of total and differential cross sections as a function of $t$ for $\gamma~p \rightarrow J/\psi~p \rightarrow (\mu^+~\mu^-)~p$~\cite{007:2026dow}. The two datasets were collected simultaneously (Table~\ref{tbl:tab1}), but they are statistically independent. The latter is the first measurement at the threshold using the $J/\psi\to \mu^+~\mu^-$ decay mode. The normalization methods for the two results differ slightly. In the case of the first publication~\cite{Duran:2022xag}, the BH reaction provided the absolute normalization (similar to GlueX~\cite{GlueX:2019mkq}). In the second publication, the total cross sections were determined directly from the differential cross sections~\cite{007:2026dow}. 

\section{J/{\ensuremath{\psi}} Threshold Measurements using the CLAS12 Spectrometer}\label{sec4}
The CLAS12 Spectrometer~\cite{Burkert:2020akg} in JLab's Experimental Hall~B has measured differential cross sections for the reaction
$\gamma^\ast~p \rightarrow J/\psi~p \rightarrow (e^+~e^-)~p$ (Table~\ref{tbl:tab1}), where $\gamma^\ast$ is used to denote quasi-real photoproduction. The experiment was conducted with a primary electron beam. Results were published for events with a squared invariant mass of the virtual photon, $Q^2$, less than $0.1~\mathrm{GeV^2}$. 
Data were collected in the fall of 2018 and the spring of 2019, covering the photon energy range $E = 8.4 - 10.5~\mathrm{GeV}$. CLAS12 reported 12 differential cross sections as a function of $t$, from which 10 total cross sections were determined as a function of $E$~\cite{Chatagnon:2026qsv}. CLAS12 likewise used the BH reaction for absolute normalization.

\section{\texorpdfstring{$J/\psi$}{J/psi}N Scattering length: Phenomenological analysis}\label{sec5}
For the evaluation of the \textit{absolute value} of J/\ensuremath{\psi}p SL, we apply the VMD approach that links the near-threshold photoproduction total cross section of $\gamma p \to V p$ and elastic scattering $V~p \to 
V~p$~\cite{Strakovsky:2014wja, Strakovsky:2021vyk}, where V = J/\ensuremath{\psi} is applicable in our case. VMD accounts for the (virtually admixed) hadronic part of the photon, through which the photon couples to hadronic matter. The applicability of the VMD model was recently explored in Ref.~\cite{Xu:2021mju} and in a previous theoretical study~\cite{Kopeliovich:2017jpy, Boreskov:1976dj}.

To avoid adding additional theoretical uncertainties, we do not: 
(i) determine the sign of SL, 
(ii) separate the Re and Im parts of SL, nor 
(iii) extract total angular momentum 1/2 and 3/2 contributions for the J/\ensuremath{\psi}p system. 

Traditionally, the total cross section, $\sigma_t$, for near-threshold binary inelastic reactions  
\begin{equation}
    a~b\to c~d
\label{eq:eq1}
\end{equation}
with 
\begin{equation}
    m_a + M_b < m_c + M_d 
\label{eq:eq2}
\end{equation}
is expressed as a series in \textit{odd} powers of the CM momentum $q$ of the V: 
\begin{equation}
    \sigma_t = a~q + b~q^3 + c~q^5
    \>,
\label{eq:eq3}
\end{equation}
where the linear term reflects the contributions of two independent $S$-wave amplitudes with a total spin of 1/2 and/or 3/2. Contributions to the cubic term originate from both $P$-wave amplitudes and the CM energy ($W$) dependence of $S$-wave amplitudes, while the fifth-order term arises from $D$-wave amplitudes as well as from the $W$ dependencies of the $S$- and $P$-wave amplitudes.
We assume that there are no VN resonance states below the experimental $q_{min}$, which are provided in Table~\ref{tbl:tab2}. 

The absolute value of the VN SL ($|\alpha_{V N}|$) is determined by the interplay of strong or hadronic ($h^2 = a$ from Eq.~(\ref{eq:eq3})) and electromagnetic (EM) dynamics ($B^2 = \frac{\alpha}{12\pi}~\frac{m_v~k_{thr}}{\Gamma(V\to e^+~e^-)}$)~\cite{Strakovsky:2014wja}:
\begin{equation}
    |\alpha_{V N}|^2~=~(h~B)^2~=~a~\biggr[\frac{\alpha}{12\pi}~\frac{M_V~k_{thr}}{\Gamma(V\to e^+~e^-)} \biggr] 
    \>,
    \label{eq:eq4}
\end{equation}
where $\alpha$ is the fine structure constant (1/137.036)~\cite{ParticleDataGroup:2024cfk}, $M_V$ is the mass of the V~\cite{ParticleDataGroup:2024cfk}, $k_{thr}$ is the photon threshold momentum in the CM for V photoproduction on the proton, $a$ is the linear coefficient obtained from fits to the V photoproduction cross sections (Eq.~(\ref{eq:eq3})), and $\Gamma(V \to e^+~e^-) = 5.53\pm 0.10~\mathrm{keV}$ is the partial decay width~\cite{ParticleDataGroup:2024cfk}. Following PDG2024, one can assume that $\Gamma(J/\psi\to e^+~e^-) = \Gamma(J/\psi\to \mu^+~\mu^-)$~\cite{ParticleDataGroup:2024cfk}. 

\begin{figure}[htb!]
\centering
{
    \includegraphics[width=0.5\textwidth,keepaspectratio]{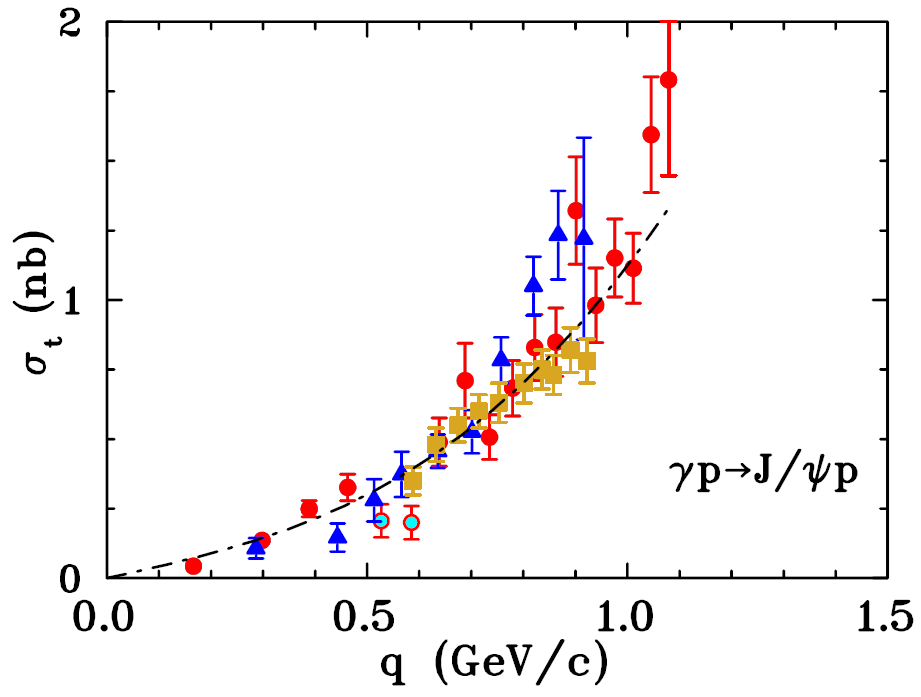}
} 
\centerline{\parbox{0.85\textwidth}{
\caption[] {\protect\small
The total $\gamma~p \to J/\psi~p$ cross sections, $\sigma_t$, as a function of the J/\ensuremath{\psi} CM momentum reported by the GlueX Collaboration (red filled circles)~\cite{GlueX:2023pev} (two points are filled in cyan at the dip around $q = 0.5 - 0.6~\mathrm{GeV/c}$ are discussed in the main text), the 007 Collaboration (shown in yellow-filled squares~\cite{007:2026dow}), and the CLAS12 Collaboration (shown in blue-filled triangles)~\cite{Chatagnon:2026qsv}.
The black dash-dotted curves represent polynomial fits to $\sigma_t$ using 
Eq.~(\ref{eq:eq3}).
} 
\label{fig:fig1} } }
\end{figure}

The fit of high-luminosity GlueX, recent 007, and CLAS12 data is shown in Fig.~\ref{fig:fig1} by a black dot-dashed curve. The best-fit results are summarized in Table~\ref{tbl:tab3}.

\begin{table}[htb!]
\centering \protect\caption{
The J/\ensuremath{\psi}-p scattering lengths extracted from the GlueX measurements on a free proton. 
The 1st column lists the minimal CM momentum $q_{min}$ for J/\ensuremath{\psi} accessible in the photoproduction experiments. 
The 2nd column gives the linear coefficient $a$ obtained from fits to the J/\ensuremath{\psi} production cross 
sections using Eq.~(\ref{eq:eq3}). 
The 3rd column gives $\chi^2/\mathrm{dof}$. 
The 4th column presents the corresponding SL, derived from the parameter $a$ via the phenomenological approach using Eq.~(\ref{eq:eq4}). 
$^\star$Fit to only the GlueX low-luminosity data.
$^\dag$Fit to only the GlueX high-luminosity data.
$^\ddag$Fit to the GlueX high-luminosity data excluding two data points ($q = 0.5 – 0.6~\mathrm{GeV/c}$) associated with a dip possibly attributable to the LHCb $P_c(4312)^+$ state~\cite{LHCb:2019kea}. 
$^\diamond$Combined fit of GlueX high-luminosity, 007, and CLAS12 data.
} 
\vspace{2mm}
{%
\begin{tabular}{|c|c|c|c|}
\hline
$q_{min}$ & $a$ & $\chi^2/\mathrm{dof}$ & $|\alpha_{J/\psi p}|$ \tabularnewline
(MeV/c) & ($\times 10^{-6}~\mathrm{\mu b/(MeV/c)}$) & & (mfm) \tabularnewline
\hline
244~\cite{GlueX:2019mkq} & 0.46$\pm$0.16~\cite{Strakovsky:2019bev}$^\star$  & 0.64 & 3.08$\pm$0.55~\cite{Strakovsky:2019bev} \tabularnewline
166~\cite{GlueX:2023pev} & 0.43$\pm$0.10~\cite{Strakovsky:2025rsm}$^\dag$ & 1.56 & 2.89$\pm$0.34~\cite{Strakovsky:2025rsm} \tabularnewline
166~\cite{GlueX:2023pev} & 0.44$\pm$0.10$^\ddag$ & 0.97 & 2.95$\pm$0.34 \tabularnewline
166~\cite{GlueX:2023pev} & 0.41$\pm$0.08$^\diamond$ & 0.93 & 2.74$\pm$0.27 \tabularnewline
\hline
\end{tabular}} \label{tbl:tab3}
\end{table}

The stability of the leading linear term $a$ from Eq.~(\ref{eq:eq3}) is demonstrated by comparing fits to the low (high) luminosity GlueX data and the combined fit of high luminosity GlueX, 007, and CLAS12 results (Table~\ref{tbl:tab3}).

One concern is the apparent dip in the GlueX data at approximately $q = 0.5 
- 0.6~\mathrm{GeV/c}$. This could hypothetically be caused by destructive 
interference between a resonant s-channel pentaquark state, plausibly the 
$P_c(4312)^+$ observed at LHCb~\cite{LHCb:2019kea}, and non-resonant J/\ensuremath{\psi} 
production~\cite{Strakovsky:2023kqu}. If this were the case, it could bias 
the results of a fit based on Eq.~\ref{eq:eq3}. Therefore, we have examined 
the results of fits to the GlueX data with ($\chi^2/dof = 1.56$) and without 
($\chi^2/dof = 0.97$) two data points at the location of the dip (indicated 
by cyan-filled points in Fig.~\ref{fig:fig1}. 
Of course, the pure polynomial expansion is not sufficient 
to describe the presence of the resonance, and accounting for these points, 
we get a larger $\chi^2$; however, as seen from Table~\ref{tbl:tab3},  thanks to the 
narrow width of the pentaquark resonance, the resulting values of SL are 
practically the same (within the error bars). That is, we see that the 
presence or the absence of the resonance does not affect our SL result.
While the 007 data lacks 
coverage in $q$ to make a firm statement about the existence of a dip, the 
more recent CLAS12 electroproduction data do not show evidence of a dip. Still, uncertainties are very large (low statistics).

In addition to examining the total cross sections, this dip may also be considered from the perspective of differential cross sections. The high-luminosity GlueX paper~\cite{GlueX:2023pev} reported $d\sigma/dt$ for three photon energy ranges: 
(i)   $E =  8.20 -  9.28~\mathrm{GeV}$, 
(ii)  $E =  9.28 - 10.36~\mathrm{GeV}$, and 
(iii) $E = 10.36 - 11.44~\mathrm{GeV}$. 
Fig.~13 of Ref.~\cite{GlueX:2023pev} demonstrated that distributions (ii) and (iii) look similar, and the data were fitted with two exponents. At the same time, distribution (i) showed that the slope of one exponent changed sign.  This bin is associated with the dip at $q = 0.5 - 0.6~\mathrm{GeV/c}$. This effect may indicate a singularity associated with the results reported in Ref.~\cite{Strakovsky:2023kqu}. 

\section{Vector Meson Nucleon Scattering Length Puzzle}\label{sec6}
In this section, we aim to summarize our current state in the VN SL puzzle.

\begin{figure}[htb!]
\centering
\includegraphics[scale=0.45]{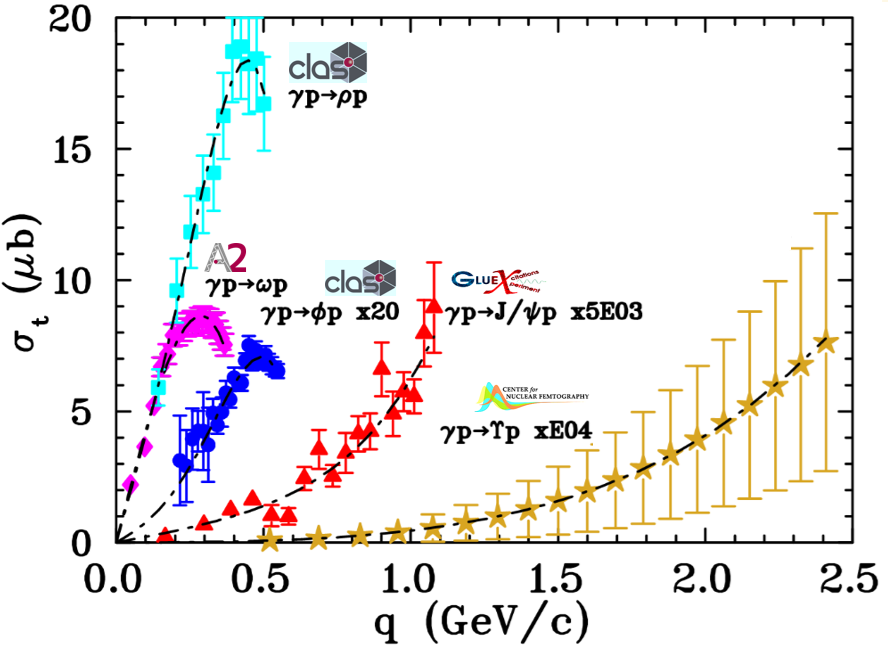}
\centerline{\parbox{0.85\textwidth}{
\caption[] {\protect\small
The total threshold $\gamma~p \to V~p$ cross section, $\sigma_t$, derived from the A2 (magenta filled diamonds)~\cite{Strakovsky:2014wja}, CLAS12 (cyan filled squares)~\cite{Strakovsky:2025ews}, CLAS6 (blue filled circles)~\cite{Strakovsky:2020uqs}, and high-luminosity GlueX (red filled triangles)~\cite{Strakovsky:2019bev} data, and simulated EIC quasi-data~\cite{Strakovsky:2021vyk} based on the theoretical predictions of Ref.~\cite{Guo:2021ibg} (based at the Center for Nuclear Femtography) are shown as a function of the CM momentum $q$ of the final-state Vs. The vertical error bars represent the total uncertainties of the data, which sum statistical and systematic uncertainties in quadrature. Black dash-dotted curves show the fits of the data and quasi-data to a series of odd powers of $q$ using Eq.~(\ref{eq:eq3}).
} 
\label{fig:fig2} } }
\end{figure}

Fits to data from A2, CLAS6, CLAS12, GlueX, and 007 (as well as to theoretical predictions in the case of $\Upsilon$~\cite{Strakovsky:2021vyk}) for the total threshold $\gamma p \to Vp$ cross sections are shown in Fig.~\ref{fig:fig2} as black dot-dashed curves.

\begin{figure}[t]
\centering
\includegraphics[scale=0.4]{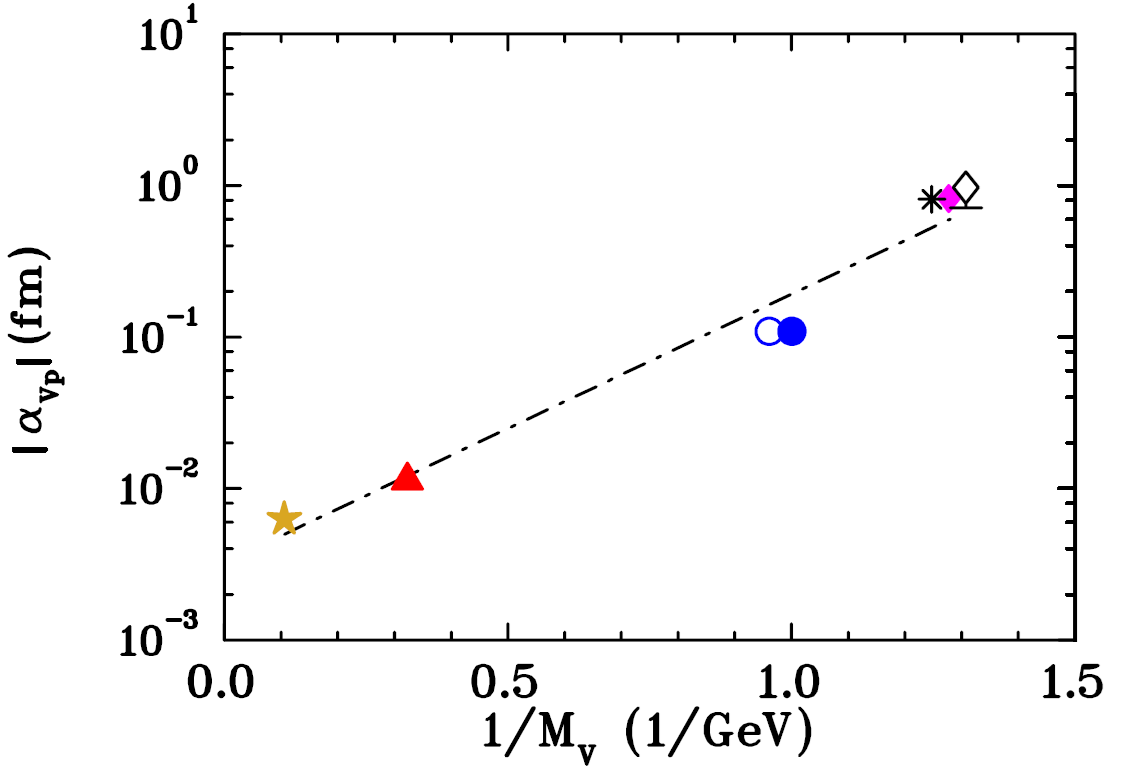}
\centerline{\parbox{0.85\textwidth}{
\caption[] {\protect\small
Comparison of $|\alpha_{Vp}|$ estimated from threshold $V$ photoproduction on the proton target assuming VMD versus the inverse mass of the $V$s, \textit{i.e.}, $1/M_V$. Input data for phenomenological analyses came from A2 at MAMI (magenta filled diamond)~\cite{Strakovsky:2014wja}, ELPH (black open diamond)~\cite{Ishikawa:2019rvz}, and CBELSA/TAPS (black asterisk)~\cite{Han:2022khg} Collaborations for the $\omega$-meson case; 
CLAS6 (blue filled circle)~\cite{Strakovsky:2020uqs} and LEPS (black open circle)~\cite{Han:2022khg} Collaborations for the $\phi$-meson case; GlueX, 007, and CLAS12 (red filled triangle) Collaborations for the J/\ensuremath{\psi}-meson case; and theoretical predictions from the Center for Nuclear Femtography (yellow filled star)~\cite{Strakovsky:2021vyk, Guo:2021ibg} for the $\Upsilon$-meson case. The black dash-dotted line is a hypothetical fit to the data from Refs.~\cite{Strakovsky:2014wja, Strakovsky:2020uqs, Strakovsky:2021vyk} with high-luminosity GlueX, 007, and CLAS12 for J/\ensuremath{\psi}p results using Eq.~(\ref{eq:eq5}) and corrected by the nucleon pole form factor.
} 
\label{fig:fig3} } }
\end{figure}

The corresponding results for the SLs are shown in Fig.~\ref{fig:fig3} as a function of the inverse vector-meson mass, $1/M_V$. Starting from the \ensuremath{\phi}-meson and for higher masses, the SLs are significantly smaller than the typical hadron size of about $1~\mathrm{fm}$, indicating an increasing transparency of the proton for these mesons. 

Note that the results in Fig.~\ref{fig:fig3} do not account for the $t$ dependence of $f_V F_\psi F_N$ form factors, while for the low energies at the threshold, the value of $t=t_{min}$ can be quite large ($t=-2.2~\mathrm{GeV^2}$ for $J/\psi$ photoproduction).

To account for the form factor, we have to divide the obtained SL by $F_N(t)$\footnote{As discussed earlier, thanks to the small size of the heavy vector meson, we take $f_V F_\psi=1$.}.

\begin{table*}[htb!]
\centering \protect\caption{The values of obtained scattering length for vector meson photoproduction neglecting, SL, and accounting,  SL/$F_N(t)$, for the nucleon form factor effects.
1st column is the name of V.
2nd column is the phenomenological SL using the conventional VMD. 
3rd column is the $t = t_{min}$.
4th column is the corrected SL using the nucleon pole form factor, $F_N(t)$, using Eq.~(\ref{3}). No theoretical uncertainties added.
5th column is the corrected SL using the nucleon exponential form factor, $b=0.76~\mathrm{GeV^{-2}}$.
6th column is the predictions using naive pQCD (current paper).
7th column is the predictions using the QCD model from Ref.~\cite{Tang:2025qqe}.
$^\dag$Joint fit result of the threshold total cross sections of the high-luminosity GlueX, 007, and CLAS12 J/$\psi$ data from the 4th raw of Table~\ref{tbl:tab3}.
Uncertainties in the 2nd column came from our phenomenological fit of total threshold cross sections of heavy vector meson photoproduction using 
Eq.~(\ref{eq:eq3}).  While uncertainties in the 4th column do not include the theoretical uncertainties.}
\vspace{2mm}
{%
\begin{tabular}{|c|c|c|c||c||c|c|}
\hline
Meson      & Photoproduction SL & $-t_{min}$ & SL/F$_N(t)$ & SL/exp(bt) & pQCD & dynSL~\cite{Tang:2025qqe} \tabularnewline
           &  (fm)          & ($\mathrm{GeV^2}$) & (fm)     & (fm)     & (fm) & (fm) \tabularnewline
\hline
$\phi$     & 0.063$\pm$0.010~\cite{Strakovsky:2020uqs} & 0.50 & 0.109$\pm$0.017 & 0.092 & 0.065 & 0.123 \tabularnewline
J/$\psi$   & 0.00274$\pm$0.00027$^\dag$ & 2.22 & 0.0113$\pm$0.0011 & 0.016 & 0.0029& 0.00634 \tabularnewline
$\Upsilon$ & 0.00051$\pm$0.00003~\cite{Strakovsky:2021vyk} & 8.07 & 0.0063$\pm$0.0004 & 0.23 & 0.00047 & 0.00196 \tabularnewline
\hline
\end{tabular}} \label{tbl:tab4}
\end{table*}

This enlarges the values for the SLs, but the results (shown in Table~\ref{tbl:tab4}) are still much less than 
$1~\mathrm{fm}$\footnote{At the moment, we have used the pole form factor (\ref{3}). Using the exponential form, $F_N=F_{exp}(t)=\exp(bt)$ produces a strange, non-physical result. Indeed, let us fix the slope $b$ from the proton electromagnetic radius, $r_e=0.84~\mathrm{fm}$. Since in the electromagnetic case the photon transfers its momentum to {\em one} quark, while in our case, via the two gluon exchange, this momentum is transferred to two quarks, we take the 4 times smaller value, $b=<r^2_e>/(6\times 4)$. This leads to the results shown in the 5th column of Table~\ref{tbl:tab4}. Note that now we get the $\Upsilon$-p SL {\em larger} than that for $J/\psi$, which is non-physical}.

\subsection{Theoretical Uncertainties}\label{sec6-A}
Aside from the experimental error bars, the uncertainties in our results arise from the accuracy of the VMD model.

First, this is the possible contribution of the quark-antiquark states (produced by the photon), other than the vector meson of interest. In terms of the dispersion relation over the mass of the intermediate state, besides the resonance peak, there may exist another contribution. The hope is that this contribution is negligibly small.

Next, the role of the form factors. On one hand, the momentum transfer needed to produce a heavy vector meson at threshold is quite large (see Table~\ref{tbl:tab4}). On the other hand, the meaning/physics of the form factor is the probability of keeping the particle intact, and not producing some complicated excited state. However, working very close to the threshold, we have no phase space sufficient to produce the new system with the same quantum numbers (say, the baryon charge B=1), except for the proton and our vector meson. That is, it seems reasonable to set $F_N$=1 (and $f_V F_\psi=1$) and say that the dynamics of interaction account for the large momentum transferred; therefore, we must consider the obtained SL (3rd column in Table~\ref{tbl:tab4}) as the real scattering length of the photoproduction on the proton.

Finally, recall that we are actually dealing with the $Q\bar Q$-proton interaction and not with the elastic $Vp\to Vp$ scattering. The size of this pair, produced by the point-like photon, is smaller than the normal vector meson size\footnote{We calculate the amplitude of $Q\bar Q\to V$ transition within the non-relativistic QCD using the vector meson wave function $\Psi(r=0)$ at zero ($r$ is the quark-quark separation).}. Thus, it is natural to expect a smaller cross-section. This ``young effect'' explains the relatively small values of SL measured in photoproduction.

\subsection{Some Speculations}\label{sec6-B}
As seen in Fig.~\ref{fig:fig3}, our analysis shows an almost linear increase (on an exponential scale) in 
\begin{equation}
    |\alpha_{V p}| \propto \exp(1/M_V)
    \>,
\label{eq:eq5a}
\end{equation}
with increasing $1/M_V$. This is an interesting observation; however, the behavior (\ref{eq:eq5a}) does not satisfy the expected asymptotics $\alpha_{Vp}\to 0$ as $M_V\to\infty$.

In fact, in pQCD, the $p\to V$ scattering amplitude is driven by the strong coupling $\alpha_s$ and the separation of the corresponding quarks. This separation (in zero approximation) is proportional to $1/M_V$, or $1/m_Q$, where $m_Q$ is the \textit{current} quark mass.

Taking the \textit{current} quark masses as $m_s = 0.1~\mathrm{GeV}$, $m_c = 1.3~\mathrm{GeV}$, and $m_b = 4.2~\mathrm{GeV}$, along with the constant $\Lambda=350~\mathrm{MeV}$~\cite{ParticleDataGroup:2024cfk}, and inserting
\begin{equation}
    \alpha_{V p} = \frac 1{30} ~ 0.197[fm\cdot GeV]~\frac{\alpha_s(m_Q)}{m_Q}
    \>,
    \label{eq:pert}
\end{equation}
with one-loop 
\begin{equation}
    \alpha_s = (4\pi/b_0)/\ln(m_Q^2/\Lambda^2) 
    \>,
    \label{eq:alphaSa}
\end{equation}
but if $\alpha_s > 1$, then $\alpha_s = 1$ (\textit{i.e.}, $\alpha_s$ is frozen at 1; and $b_0$ is the one-loop $\beta$-function coefficient $b_0 = 8.33$ for 4 light quarks: $u$, $d$, $s$, and $c$). 
One can obtain rather good results: 
$|\alpha_{\phi N}| = 65~\mathrm{mfm}$, 
$|\alpha_{J/\psi N}| = 2.9~\mathrm{mfm}$, and 
$|\alpha_{\Upsilon N}| = 0.47~\mathrm{mfm}$.

Of course,  the expression (\ref{eq:pert}) is not the real pQCD calculation. This is just a naive estimate. However, it demonstrates that the observed SLs are consistent with pQCD and account for the suppression due to the young effect.

Recall that the aim of our paper is not to build a model that describes the data, but rather to conduct a phenomenological analysis. (The model(s) and the corresponding theoretical uncertainties can be found, \textit{e.g.}, in~\cite{Tang:2025qqe, Du:2020bqj}.)

\subsection{Other Comments}\label{sec6-C}
Actually, Eq.~(8) from the paper by Titov \textit{et al.}~\cite{Titov:2007xb} gives the relationship between the scattering lengths of \ensuremath{\phi}N and \ensuremath{\omega}N using $SU(3)_{flavor}$ symmetry:
\begin{equation}
    |\alpha_{\phi N}|~=~\epsilon~|\alpha_{\omega N}| 
    \>,
    \label{eq:eq6}
\end{equation}
where $\epsilon \equiv -tg(\Delta\theta_V)$ and $\Delta\theta_V \simeq 3.7^\circ$ represents the deviation of the \ensuremath{\phi} - \ensuremath{\omega} mixing angle from the ideal mixing angle. (The vector mixing angle $\theta_V = 36.5^\circ$ is very close to the ideal mixing angle~\cite{ParticleDataGroup:2024cfk}.) The relation (\ref{eq:eq6}) assumes that SL originates from the light ($u$, $d$) quarks only, most likely from the nucleon resonances. The ratio of our phenomenological SL determination 
($|\alpha_{\omega N}| = 0.82~\mathrm{fm}$~\cite{Strakovsky:2014wja} and $|\alpha_{\phi N}| = 0.063~\mathrm{fm}$~\cite{Strakovsky:2020uqs}), 
taken from Table~\ref{tbl:tab2}, is
$\epsilon = \frac{|\alpha_{\phi N}|}{|\alpha_{\omega N}|} = \frac{0.063}{0.82} = 0.077$. It corresponds to $\Delta\theta_v \simeq 4.4^\circ$. That is in perfect agreement with the result from~\cite{Titov:2007xb}.

In $SU(3)_{flavor}$, there should be a similar relation for 
$|\alpha_{\omega N}|$ with $|\alpha_{\rho N}|$ and $|\alpha_{\phi N}|$ with $|\alpha_{\rho N}|$. However, for the \ensuremath{\rho}, it may be partly spoiled by the large \ensuremath{\rho} width. Besides this, in the \ensuremath{\rho}p case, the result may be affected by the {\em isobars} ($I=3/2$). That is likely why $|\alpha_{\omega N}| \simeq 
4~|\alpha_{\rho N}|$, as was reported recently~\cite{Strakovsky:2025ews}.\\

\section{J/\ensuremath{\psi} Measurements using MARQ and Dilepton Detectors at 
J-PARC}\label{sec7}
The new high-energy \ensuremath{\pi}20 beamline, utilized by the P111 experiment~\cite{Ryu:2024}, allows precise threshold measurements of the $\pi^-~p \to {\rm J}/\psi~n$ reaction using a method independent of the VMD model. The J/\ensuremath{\psi} production will be reconstructed in both $e^+~e^-$ and $\mu^+~\mu^-$ di-lepton decay channels. The near-threshold cross section for the $\pi^-~p \to {\rm J}/\psi~n$ reaction remains unmeasured; previous studies have only established upper limits. Theoretical predictions for the J/\ensuremath{\psi} production cross section span a wide range, from $0.1~\mathrm{pb}$ to $50~\mathrm{nb}$~\cite{Sibirtsev:1998cs, Wu:2013xma, Lu:2015fva, Kim:2016cxr}, depending on interpretations of these limits. If the cross section reaches approximately $4~\mathrm{nb}$ at $p = 9.5~\mathrm{GeV/c}$, the anticipated J/\ensuremath{\psi} yield in the momentum range $p = 8.2 - 
11.0~\mathrm{GeV}/c$ is estimated to be $1.0\times 10^4~\mathrm{events}$. These results are essential for advancing the phenomenological understanding of the SL puzzle (in particular, the role of form factors near the threshold) and for investigating potential effects related to the LHCb heavy pentaquarks ($P_c$)~\cite{LHCb:2019kea}.

Furthermore, the J-PARC experiment aims to search for isospin partners of the $P_c$ states via the $\pi^-~p \to {\rm J}/\psi~n$ reaction near the threshold. The P111 setup, which integrates both the MARQ and di-lepton detectors, enables precise measurements of both hidden-charm and open-charm channels - including $\Lambda_c^{(\ast)}\bar{D}^{(\ast)}$ and $\Sigma_c^{(\ast)}\bar{D}^{(\ast)}$ - thus allowing a comprehensive coupled-channel analysis near the $P_c$ states~\cite{LHCb:2019kea}. If the $P_c$ states strongly couple to $\Sigma_c^{(\ast)}\bar{D}^{(\ast)}$, the pion-induced reaction would be expected to dominate over the photon-induced reaction, provided $g_{\Lambda_cN\bar{D}} \gg g_{\Sigma_cN\bar{D}}$. 

\section{Summary}\label{sec8}
The J/\ensuremath{\psi}-proton SL, phenomenologically determined using recent 007 and CLAS12 measurements of J/\ensuremath{\psi} photo(electro)production at threshold using both $J/\psi \to l^+l^-$ decay channels, together with previous GlueX measurements with the $J/\psi \to e^+e^-$ decay, is in perfect agreement. 

We used the concept of the young vector meson hypothesis, which may explain the fact that the obtained SL value for \ensuremath{\phi}-meson nucleon, compared to the typical hadron size of $1~\mathrm{fm}$, indicates that the proton is more transparent for \ensuremath{\phi}-meson than for \ensuremath{\omega}-meson and is much more transparent for J/\ensuremath{\psi}-meson: 
\begin{equation}
    |\alpha_{\Upsilon p}| \ll |\alpha_{J/\psi p}| \ll |\alpha_{\phi p}| \ll |\alpha_{\omega p}|
    \>.
\label{eq:eq7}
\end{equation}

We have shown that the pQCD estimate is consistent with our phenomenological treatment of the heavy-vector-meson photoproduction data at threshold.

The puzzle of the dynamics of $q\bar{q}$, $s\bar{s}$, and $c\bar{c}$ interactions with the $qqq$ system at the threshold remains an open question that the upcoming J-PARC measurements will resolve.

Polarized measurements are essential for model-independent partial-wave analysis (PWA). J/\ensuremath{\psi} is somewhat polarized, but it is much less polarized than Drell-Yan di-leptons, as FNAL measurements of the polarization of J/\ensuremath{\psi} in $pA$ inclusive production at $800~\mathrm{GeV}$ have shown~\cite{NuSea:2000vgl, NuSea:2003fkm}. The J/\ensuremath{\psi} polarization matrix is measured via the angular distribution of electrons in $J/\psi \to e^+~e^-$ decay. The JLab data and J-PARC experiment P111 may help resolve the polarization effects in quarkonium production. 

\section{Acknowledgments}\label{sec9}
We thank our Referee for discussing the nucleon form factor contribution.
We thank Yuri Dokshitzer, Atsushi Hosaka, Sasha Titov, Jen Chieh Peng, Pierre Chatagnon, Sylvester Joosten, Valery Kubarovsky, and Sun Young Ryu for their valuable comments and discussions. This work was supported in part by the U.S.~Department of Energy, Office of Science, Office of Nuclear Physics, under Award No. DE--SC0016583, and by the National Research Foundation of Korea (NRF), the Korean government (MSIT), under Grant No. RS--2024--00436392.



\begin{thebibliography}{99}
\bibitem{Sakurai:1960ju}
    J.~J.~Sakurai,
    ``Theory of strong interactions,''
    Ann.\ Phys.\ (N.Y.) \textbf{11}, 1 (1960).
\bibitem{E598:1974sol}
    J.~J.~Aubert \textit{et al.} [E598 Collaboration],
    ``Experimental observation of a heavy particle $J$,''
    Phys.\ Rev.\ Lett.\ \textbf{33}, 1404 (1974).
\bibitem{SLAC-SP-017:1974ind}
    J.~E.~Augustin \textit{et al.} [SLAC-SP-017 Collaboration],
    ``Discovery of a narrow resonance in $e^+~e^-$ annihilation,''
    Phys.\ Rev.\ Lett.\ \textbf{33}, 1406 (1974).
\bibitem{GlueX:2019mkq}
    A.~Ali \textit{et al.} [GlueX Collaboration],
    ``First measurement of near-threshold J/\ensuremath{\psi} exclusive photoproduction off the proton,''
    Phys.\ Rev.\ Lett.\ \textbf{123}, 072001 (2019).
\bibitem{GlueX:2023pev}
    S.~Adhikari \textit{et al.} [GlueX Collaboration],
    ``Measurement of the J/\ensuremath{\psi} photoproduction cross section over the full near-threshold kinematic region,''
    Phys.\ Rev.\ C\ \textbf{108}, 025201 (2023).
\bibitem{Duran:2022xag}
    B.~Duran, Z.~E.~Meziani, S.~Joosten, M.~K.~Jones, S.~Prasad, C.~Peng, W.~Armstrong, H.~Atac, E.~Chudakov, H.~Bhatt \textit{et al.}
    ``Determining the gluonic gravitational form factors of the proton,''
    Nature\ \textbf{615}, 813 (2023).
\bibitem{007:2026dow}
    S.~Joosten \textit{et al.} [007 Collaboration],
    ``Near-threshold J/$\psi\to \mu^+\mu^-$ photoproduction and the Gluonic Gravitational Form Factors of the proton,''
    [arXiv:2602.14416 [nucl-ex]].
\bibitem{Chatagnon:2026qsv}
    P.~Chatagnon, V.~Kubarovsky, R.~Paremuzyan, S.~Stepanyan, M.~Tenorio, R.~Tyson, A.~G.~Acar, P.~Achenbach, J.~S.~Alvarado, M.~J.~Amaryan \textit{et al.}
    ``Measurement of the near-threshold J/\ensuremath{\psi} photoproduction cross section with the CLAS12 experiment,''
    [arXiv:2602.22128 [hep-ex]].
\bibitem{Strakovsky:2025ews}
    I.~I.~Strakovsky, E.~L.~Isupov, V.~Mokeev, and A.~Schmidt,
    ``\ensuremath{\rho}-meson nucleon scattering length from CLAS threshold photoproduction measurements,''
    Phys.\ Rev.\ D\ \textbf{112}, 114045 (2025).
\bibitem{Wang:2022zwz}
    X.~Y.~Wang, F.~Zeng, Q.~Wang, and L.~Zhang,
    ``First extraction of the proton mass radius and scattering length $\vert\alpha_{\rho^{0}p}\vert$ from {\ensuremath{\rho}}$^{0}$ photoproduction,''
    Sci.\ China\ Phys.\ Mech.\ Astron.\ \textbf{66}, 232012 (2023).
\bibitem{Strakovsky:2014wja}
    I.~I.~Strakovsky \textit{et al.} [A2 Collaboration at MAMI],
    ``Photoproduction of the \ensuremath{\omega} meson on the proton near threshold,''
    Phys.\ Rev.\ C\ \textbf{91}, 045207 (2015).
\bibitem{Ishikawa:2019rvz}
    T.~Ishikawa, H.~Fujimura, H.~Fukasawa, R.~Hashimoto, Q.~He, Y.~Honda, A.~Hosaka, T.~Iwata, S.~Kaida, J.~Kasagi \textit{et al.},
    ``\ensuremath{\omega}N scattering length from $\omega$ photoproduction on the proton near the threshold,''
    Phys.\ Rev.\ C\ \textbf{101}, 052201 (2020).
\bibitem{Han:2022khg}
    C.~Han, W.~Kou, R.~Wang, and X.~Chen,
    ``Extraction of \ensuremath{\omega}n, \ensuremath{\omega}p, and \ensuremath{\phi}N scattering lengths from \ensuremath{\omega} and \ensuremath{\phi} differential photoproduction cross sections~on a deuterium target,''
    Phys.\ Rev.\ C\ \textbf{107}, 015204 (2023).
\bibitem{Strakovsky:2020uqs}
    I.~I.~Strakovsky, L.~Pentchev, and A.~Titov,
    ``Comparative analysis of \ensuremath{\omega}p, \ensuremath{\phi}p, and J/\ensuremath{\psi}p scattering lengths from A2, CLAS, and GlueX threshold measurements,''
    Phys.\ Rev.\ C\ \textbf{101}, 045201 (2020).
\bibitem{Strakovsky:2019bev}
    I.~Strakovsky, D.~Epifanov, and L.~Pentchev,
    ``J/{\ensuremath{\psi}}p scattering length from GlueX threshold measurements,''
    Phys.\ Rev.\ C\ \textbf{101}, 042201 (2020).
\bibitem{Strakovsky:2025rsm}
    I.~I.~Strakovsky, W.~J.~Briscoe, P.~Gubler, J.~R.~Pybus, A.~Schmidt, and A.~Somov,
    ``J/\ensuremath{\psi}-meson nucleon scattering length from threshold photoproduction on light nuclei,''
    [arXiv:2512.08701 [hep-ph]].
\bibitem{Strakovsky:2021vyk}
    I.~I.~Strakovsky, W.~J.~Briscoe, L.~Pentchev, and A.~Schmidt,
    ``Threshold \ensuremath{\Upsilon}-meson photoproduction at the EIC and EicC,''
    Phys.\ Rev.\ D\ \textbf{104}, 074028 (2021).
\bibitem{ParticleDataGroup:2024cfk}
    S.~Navas \textit{et al.} [Particle Data Group],
    ``Review of particle physics,''
    Phys.\ Rev.\ D\ \textbf{110}, 030001 (2024).
\bibitem{CLAS:2018drk}
    E.~Golovatch \textit{et al.} [CLAS Collaboration],
    ``First results on nucleon resonance photocouplings from the $\gamma~p \to \pi^+~\pi^-~p$ reaction,''
    Phys.\ Lett.\ B\ \textbf{788}, 371 (2019).  
\bibitem{Klein:1996}
    F.~J.~Klein \textit{et al.} [SAPHIR Collaboration], Proceedings of the GW/TJNAF Workshop on $N^\ast$ Physics, (Washington, DC), (1997); 
    F.~J.~Klein, Ph.~D. Thesis, University of Bonn (1996).
\bibitem{CBELSATAPS:2015wwn}
    F.~Dietz \textit{et al.} [CBELSA/TAPS Collaboration],
    ``Photoproduction of \ensuremath{\omega} mesons off protons and neutrons,''
    Eur.\ Phys.\ J.\ A\ \textbf{51}, 6 (2015).
\bibitem{Dey:2014tfa}
    B.~Dey \textit{et al.} [CLAS Collaboration],
    ``Data analysis techniques, differential cross sections, and spin density matrix elements for the reaction $\gamma~p \rightarrow \phi~p$,''
    Phys.\ Rev.\ C\ \textbf{89}, 055208 (2014).
\bibitem{Chang:2007fc}
    W.~C.~Chang, K.~Horie, S.~Shimizu, M.~Miyabe, D.~S.~Ahn, J.~K.~Ahn, H.~Akimune, Y.~Asano, S.~Date, H.~Ejiri \textit{et al.}
    ``Forward coherent \ensuremath{\phi}-meson photoproduction from deuterons near threshold,''
    Phys.\ Lett.\ B\ \textbf{658}, 209 (2008);
    T.~Mibe \textit{et al.} [LEPS Collaboration],
    ``Diffractive \ensuremath{\phi}-meson photoproduction on proton near threshold,''
    Phys.\ Rev.\ Lett.\ \textbf{95}, 182001 (2005).
\bibitem{GlueX:2023fcq}
    S.~Adhikari \textit{et al.} [GlueX Collaboration],
    ``Measurement of spin-density matrix elements in {\ensuremath{\rho}}(770) production with a linearly polarized photon beam at E{\ensuremath{\gamma}}=8.2{\textendash}8.8~GeV,''
    Phys.\ Rev.\ C\ \textbf{108}, 055204 (2023).
\bibitem{Guo:2021ibg}
    Y.~Guo, X.~Ji, and Y.~Liu,
    ``QCD analysis of near-threshold photon-proton production of heavy quarkonium,''
    Phys.\ Rev.\ D\ \textbf{103}, 096010 (2021).
\bibitem{Feinberg:1980yu}
    E.~L.~Feinberg,
    ``Hadron clusters and half-dressed particles in Quantum Field Theory,''
    Usp.\ Fiz.\ Nauk\ \textbf{132}, 225 (1980) 
    [Sov.\ Phys.\ Usp.\ \textbf{23}, 629 (1980)].
\bibitem{Tang:2025qqe}
    L.~Tang, H.~Y.~Xing, M.~Ding, and C.~D.~Roberts,
    ``Exclusive photoproduction of light and heavy vector mesons: thresholds to very high energies,''
    Eur.\ Phys.\ J.\ C\ \textbf{86}, 284 (2026).
\bibitem{Du:2020bqj}
    M.~L.~Du, V.~Baru, F.~K.~Guo, C.~Hanhart, U.~G.~Mei{\ss}ner, A.~Nefediev, and I.~Strakovsky,
    ``Deciphering the mechanism of near-threshold J/\ensuremath{\psi} photoproduction,''
    Eur.\ Phys.\ J.\ C\ \textbf{80}, 1053 (2020).
\bibitem{Wu:2024xwy}
    B.~Wu, X.~K.~Dong, M.~L.~Du, F.~K.~Guo, and B.~S.~Zou,
    ``Deciphering the mechanism of J/{\ensuremath{\psi}}-nucleon scattering,''
    Fund.\ Res.\ \textbf{5}, 2530 (2025).
\bibitem{Lyu:2024ttm}
    Y.~Lyu, T.~Doi, T.~Hatsuda, and T.~Sugiura,
    ``Nucleon-charmonium interactions from lattice QCD,''
    Phys.\ Lett.\ B\ \textbf{860}, 139178 (2025).
\bibitem{Appelquist:1978rt}
    T.~Appelquist and W.~Fischler,
    ``Some remarks on Van der Waals forces in {QCD},''
    Phys.\ Lett.\ B\ \textbf{77}, 405 (1978).
\bibitem{Peskin:1979va}
    M.~E.~Peskin,
    ``Short distance analysis for heavy quark systems. 1. Diagrammatics,''
    Nucl.\ Phys.\ B\ \textbf{156}, 365 (1979).
\bibitem{Liu:2008rza}
    L.~Liu, H.~W.~Lin, and K.~Orginos,
    ``Charmed Hadron Interactions,''
    PoS\ \textbf{LATTICE2008}, 112 (2008).
\bibitem{Aoki:2009ji}
    S.~Aoki, T.~Hatsuda, and N.~Ishii,
    ``Theoretical foundation of the nuclear force in QCD and its applications to central and tensor forces in quenched Lattice QCD simulations,''
    Prog.\ Theor.\ Phys.\ \textbf{123}, 89 (2010).
\bibitem{Assi:2025ysr}
    B.~Assi, A.~V.~Grebe, and M.~L.~Wagman,
    ``Baryon-baryon, meson-meson, and meson-baryon interactions in nonrelativistic QCD,''
    Phys.\ Rev.\ D\ \textbf{113}, 016013 (2026).
\bibitem{Dokshitzer:2024}
    Yu.~L.~Dokshitzer (private communication).
\bibitem{Ryu:2024}
    S.~Y.~Ryu \textit{et al.} [J-PARC P111 Collaboration],
    ``J/\ensuremath{\psi} production in $\pi^-p$ reaction near threshold,'' Proposal for Nuclear and Particle Physics Experiments at J-PARC (2024). 
\bibitem{GlueX:2020idb}
    S.~Adhikari \textit{et al.} [GlueX Collaboration],
    ``The GlueX beamline and detector,''
    Nucl.\ Instrum.\ Meth.\ A\ \textbf{987}, 164807 (2021).
\bibitem{Ali:2025dan}
    S.~Ali, A.~Ahmidouch, G.~R.~Ambrose, A.~Asaturyan, C.~Ayerbe Gayoso, J.~Benesch, V.~Berdnikov, H.~Bhatt, D.~Bhetuwal, D.~Biswas \textit{et al.}
    ``The SHMS $11~\mathrm{GeV/c}$ spectrometer in Hall~C at Jefferson Lab,''
    Nucl.\ Instrum.\ Meth.\ A\ \textbf{1083}, 171070 (2026).
\bibitem{Burkert:2020akg}
    V.~D.~Burkert, L.~Elouadrhiri, K.~P.~Adhikari, S.~Adhikari, M.~J.~Amaryan, D.~Anderson, G.~Angelini, M.~Antonioli, H.~Atac, S.~Aune \textit{et al.}
    ``The CLAS12 Spectrometer at Jefferson Laboratory,''
    Nucl.\ Instrum.\ Meth.\ A\ \textbf{959}, 163419 (2020).
\bibitem{Xu:2021mju}
    Y.~Z.~Xu, S.~Chen, Z.~Q.~Yao, D.~Binosi, Z.~F.~Cui, and C.~D.~Roberts,
    ``Vector-meson production and vector meson dominance,''
    Eur.\ Phys.\ J.\ C\ \textbf{81}, 895 (2021).
\bibitem{Kopeliovich:2017jpy}
    B.~Z.~Kopeliovich, I.~Schmidt, and M.~Siddikov,
    ``Suppression versus enhancement of heavy quarkonia in pA collisions,''
    Phys.\ Rev.\ C\ \textbf{95}, 065203 (2017).
\bibitem{Boreskov:1976dj}
    K.~G.~Boreskov and B.~L.~Ioffe,
    ``$J/\psi$ meson photoproduction in the peripheral model,''
    Yad.\ Fiz.\ \textbf{25}, 806 (1977)
    [Sov.\ J.\ Nucl.\ Phys.\ \textbf{25}, 331 (1977)].
\bibitem{LHCb:2019kea}
    R.~Aaij \textit{et al.} [LHCb Collaboration],
    ``Observation of a narrow pentaquark state, $P_c(4312)^+$, and of two-peak structure of the $P_c(4450)^+$,''
    Phys.\ Rev.\ Lett.\ \textbf{122}, 222001 (2019).
\bibitem{Strakovsky:2023kqu}
    I.~Strakovsky, W.~J.~Briscoe, E.~Chudakov, I.~Larin, L.~Pentchev, A.~Schmidt, and R.~L.~Workman,
    ``Plausibility of the LHCb $P_c(4312)^+$ in the GlueX {\ensuremath{\gamma}}~p~{\textrightarrow}~J/{\ensuremath{\psi}}~p total cross sections,''
    Phys.\ Rev.\ C\ \textbf{108}, 015202 (2023).
\bibitem{Titov:2007xb}
    A.~I.~Titov, T.~Nakano, S.~Date, and Y.~Ohashi,
    ``Comments on differential cross-section of \ensuremath{\phi}-meson photoproduction at threshold,''
    Phys.\ Rev.\ C\ \textbf{76}, 048202 (2007).
\bibitem{Sibirtsev:1998cs}
    A.~Sibirtsev and K.~Tsushima,
    ``J/\ensuremath{\phi} production in $\pi$N collisions,''
    [arXiv:nucl-th/9810029 [nucl-th]].
\bibitem{Wu:2013xma}
    J.~J.~Wu and T.~S.~H.~Lee,
    ``Production of J/\ensuremath{\phi} on the nucleon and on deuteron targets,''
    Phys.\ Rev.\ C\ \textbf{88}, 015205 (2013).
\bibitem{Lu:2015fva}
    Q.~F.~L{\"u}, X.~Y.~Wang, J.~J.~Xie, X.~R.~Chen, and Y.~B.~Dong,
    ``Neutral hidden charm pentaquark states $P_c^0(4380)$ and $P_c^0(4450)$ in $\pi^-~p \to J/\psi~n$ reaction,''
    Phys.\ Rev.\ D\ \textbf{93}, 034009 (2016).
\bibitem{Kim:2016cxr}
    S.~H.~Kim, H.~C.~Kim, and A.~Hosaka,
    ``Heavy pentaquark states $P_c(4380)$ and $P_c(4450)$ in the $J/\psi$ production induced by pion beams off the nucleon,''
    Phys.\ Lett.\ B\ \textbf{763}, 358 (2016).      
\bibitem{NuSea:2000vgl}
    C.~N.~Brown \textit{et al.} [NuSea Collaboration],
    ``Observation of polarization in bottomonium production at $\sqrt{s} = 38.8~\mathrm{GeV}$,''
    Phys.\ Rev.\ Lett.\ \textbf{86}, 2529 (2001).
\bibitem{NuSea:2003fkm}
    T.~H.~Chang \textit{et al.} [NuSea Collaboration],
    ``J/\ensuremath{\psi} polarization in $800~\mathrm{GeV}$ pCu interactions,''
    Phys.\ Rev.\ Lett.\ \textbf{91}, 211801 (2003).
\end{thebibliography}
\end{document}